\documentclass[11pt, oneside]{article}   	
\usepackage[margin=1in]{geometry}                		
\usepackage[parfill]{parskip}    		
\usepackage{graphicx}				
\usepackage{authblk}									
\usepackage{hyperref}
\usepackage{amssymb}
\usepackage{cite}
\usepackage{float}
\usepackage{fixltx2e}
\usepackage{placeins}
\usepackage{verbatim}
\usepackage{placeins}

\title{Effects of applied strain on radiation damage generation in body-centered cubic iron}

\date{}							

\begin{document}

\author[1,2]{Benjamin Beeler}
\author[2]{Mark Asta}
\author[3]{Peter Hosemann}
\author[1,4]{Niels Gr{\o}nbech-Jensen}
\affil[1]{\small\it{Department of Mechanical and Aeronautical Engineering, University of California, Davis, CA 95616}}
\affil[2]{\small\it{Department of Materials Science, University of California, Berkeley, CA 94720}}
\affil[3]{\small\it{Department of Nuclear Engineering, University of California, Berkeley, CA 94720}}
\affil[4]{\small\it{Department of Mathematics, University of California, Davis, CA 95616}}

\maketitle

\section{Abstract}

Radiation damage in body-centered cubic (BCC) Fe has been extensively studied by computer simulations to quantify effects of temperature, impinging particle energy, and the presence of extrinsic particles.  However, limited investigation has been conducted into the effects of mechanical stresses and strain.  In a reactor environment, structural materials are often mechanically strained, and an expanded understanding of how this strain affects the generation of defects may be important for predicting microstructural evolution and damage accumulation under such conditions.  In this study, we have performed molecular dynamics simulations in which various types of homogeneous strains are applied to BCC Fe and the effect on defect generation is examined.  It is found that volume-conserving shear strains yield no statistically significant variations in the stable number of defects created via cascades in BCC Fe.  However, strains that result in volume changes are found to produce significant effects on defect generation.

\section{Introduction}

Understanding materials response to irradiation is one of the key issues for the design and operation of nuclear fission and fusion power systems.  Lifetime and operational restrictions are typically influenced by limitations associated with the degradation of structural materials properties under radiation.  The ability to accurately predict and understand behavior of structural materials under short term and long term radiation damage is thus important for designing improved materials for nuclear power applications.  Extensive computational research has been performed in order to better understand radiation damage in widely used structural materials, including ferritic alloys based on body-centered cubic (BCC) iron (Fe) \cite{bacon1993, stoller1996, gao1996, gao1997, becquart1997, stoller1999, malerba2006}.  

Common areas of interest with regards to radiation damage simulations have been general damage behavior \cite{averback1998, phythian1995, bacon1994}, variations in primary knock-on atom (PKA) energy \cite{caturla2000, zinkle1993}, variations in simulation temperature during irradiation \cite{gao1997, phythian1995, soneda1998, bacon2004} and the behavior and effect of extrinsic particles \cite{hayward2010, ackland2004, terentyev2006}.  However, one topic that has received much less attention is the effect of strain on the generation of radiation damage.  Although minimally studied, it is of critical importance to understand the effect of strains on damage generation and accumulation.  Structural materials in nuclear fission and fusion systems are often exposed to applied stresses, and they are subject to a dynamic strain environment due to a variety of phenomena such as void swelling, solute precipitation, solute segregation, etc. \cite{brailsfo1972, auger2000, johnson1976}.  Thus, there is a distinct need for expanded information related to radiation damage behavior under applied strain.  

Three previous computational studies have been performed to examine strain effects on damage generation.  Miyashiro, \textit{et al.} \cite{miyashiro2011} investigated the effects of strain on damage generation in face-centered cubic (FCC) copper.  Strain types analyzed included uniaxial tension and compression along the [111] axis, hydrostatic strain, and isometric strain (tetragonal shear) with maximum strain of up to 1$\%$.  For the PKA, a random set of directions was chosen, each with an energy of 10 keV.  It was found that defect production increased with both uniaxial tension and compression.  It was also found that the largest increase in defect production was due to isometric strain.  Another study was conducted by Di, \textit{et al.} \cite{di2013} on hexagonal close-packed (HCP) zirconium.  They enforced tensile and compressive strains along the \textit{a} and \textit{c} axes with magnitudes up to 1 $\%$ strain.  Various PKA directions were investigated with energies of 10 keV.  In this study, it was concluded that the main effect of strain is on the size of the defect clusters.  However, there was no significant effect of applied strain on the total number of defects generated.  Finally, Gao, \textit{et al.} \cite{gao2001} investigated BCC Fe.  The only strain type analyzed was tension along the [111] axis with a maximum magnitude of 1 $\%$ strain.  The PKA direction was restricted to the [135] direction with an energy of 10 keV.  It was found that there was a slight reduction in the point defect generation for 0.1 $\%$ strain.  However, higher strains seemed to yield negligible changes versus the unstrained system.  

These previous works have provided important insights into the effects of elastic strain on radiation damage production.  However, particularly for BCC Fe, the computational investigations are incomplete.  Gao, \textit{et al.} \cite{gao2001} investigated one type of strain and one direction for the PKA.  To develop a more detailed picture of the effects of strain on radiation damage, various types of applied strain are analyzed for various PKA directions in the current work.  Specifically, we apply molecular dynamics (MD) \cite{berendsen1984} simulations of pure BCC Fe under several types of applied strain, to analyze the resulting effects on damage generation.

\section{Computational Details}
Molecular dynamics simulations are performed utilizing the LAMMPS \cite{plimpton1995} software package and the Embedded-Atom Method (EAM) \cite{daw1984} interatomic potential developed for BCC Fe by Mendelev \textit{et al.} \cite{mendelev2003}.  A BCC supercell containing 250,000 atoms is relaxed for 100 ps at 300 K in an NPT ensemble.  A strain is applied to the supercell and the strained lattice is allowed to relax for 100 ps in an NVT ensemble.  An atom near the center of the supercell is then given extra kinetic energy, with the velocity directed in varying prescribed directions.  The time step is set to 0.2 fs and the simulation is run for 75000 steps.  We utilize the GJF thermostat \cite{gjf2013, gjf2014} (fix langevin gjf) due to its robust configurational sampling properties.  The damping parameter (analogous to relaxation time) is set to 1 ps.  The number of stable Frenkel pairs is determined after 15 ps via a Wigner-Seitz cell based algorithm \cite{hayward2010}.  Thus, this analysis does not take into account long-time thermal diffusion.  All PKA energies are set to 5 keV.  The PKA directions analyzed include [135], [100], [001], [110], and finally a set of 32 randomly selected directions.  For each strain state and PKA direction, 32 independent simulations are performed with a unique distribution of initial velocities.  

We have chosen the PKA directions in order to provide a representative set of data for the BCC crystal.  High symmetry directions parallel to and perpendicular to the applied strain could potentially see different effects based on the distortion of the local environment.  Also, the majority of PKA collisions will not be along high symmetry directions.  In the literature involving radiation damage simulations, the [135] direction is commonly utilized as a direction representing average behavior for the BCC crystal system \cite{stoller1997}.  This direction also has the added benefit of reducing channelling.  Thus, we analyze the [135] direction, as well as a set of 32 randomly chosen directions, allowing us to compare a direction commonly accepted as representative of random average behavior, to a set of truly random directions.  The random directions were created by varying the tilt $\theta$ and azimuthal $\phi$ angles displayed in Figure 1.  $\theta$ values were selected in the range 0 to $\pi$/2, $\phi$ values were selected in the range 0 to $\pi$/4.   These ranges of $\theta$ and $\phi$ encompass a complete sampling of directions in a BCC supercell.  

\begin{figure}[h!]
   \centering
   \includegraphics[width=\textwidth]{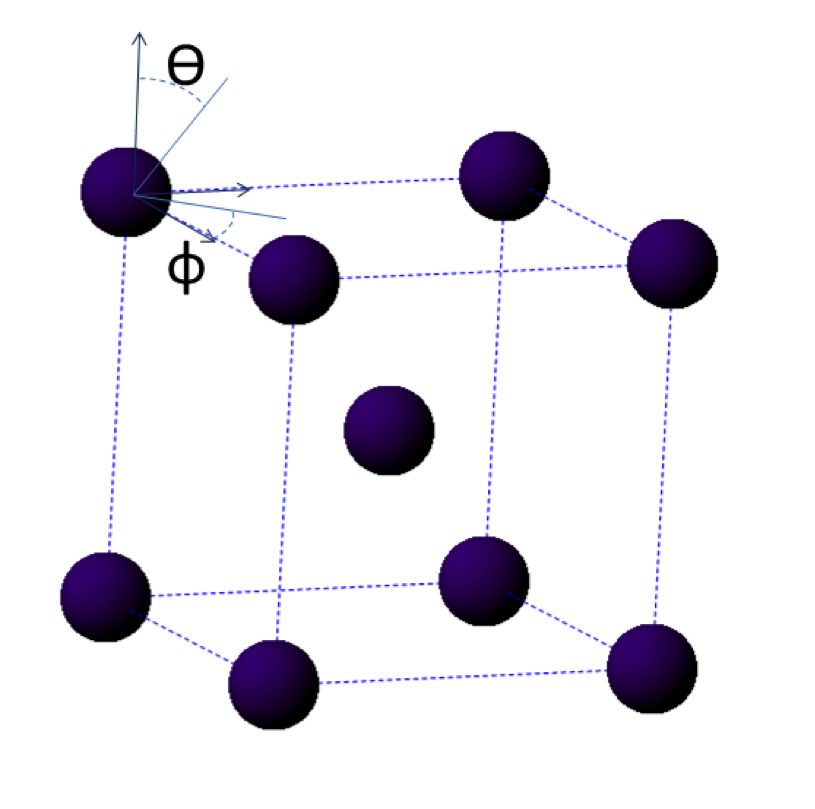} 
   \caption{Random directions were selected for primary knock-on atoms to sample the complete set of directions within a BCC crystal system.  The angle theta ($\theta$) was varied from 0 to $\pi$/2.  The angle phi ($\phi$) was varied from 0 to $\pi$/4.}
   \label{fig:example}
\end{figure}

\section{Results}

The number of defects created under applied strain for isotropic expansion/compression (hydrostatic strain), uniaxial strain, monoclinic shear and tetragonal shear strains are displayed in Figure 2.  Details of each deformation type are shown in Table 1 for a representative 1$\%$ applied strain. For hydrostatic and uniaxial strains, results for compressive (negative) and tensile (positive) strains of up to 2$\%$ are displayed in Figure 2.  Monoclinic shear strains are investigated up to 2$\%$ and tetragonal shear strains are investigated up to 5$\%$.  Error bars included in Figure 2 denote twice the standard error of the mean, which is defined as the standard deviation divided by the square root of the sample size.

Figure 2\textit{a} depicts results for hydrostatic strain, with PKA directions along [135], [100], [110] and average behavior generated by averaging results over 32 random orientations (denoted henceforth simply as random directions).  For all directions analyzed, there appears to be significant variance with applied expansion and compression.  As the system expands, there is a consistent trend towards increased defect generation.  As the system is compressed, there is a consistent trend towards decreased defect generation.  

\begin{table}[htbp]
\caption{Details of each deformation for a representative 1$\%$ applied strain.}
\begin{center}
\begin{tabular}{|c|c|c|c|c|c|}
	\hline
	& $\epsilon$$_{xx}$ $\%$ & $\epsilon$$_{yy}$ $\%$ & $\epsilon$$_{zz}$ $\%$ & $\epsilon$$_{xy}$ $\%$ & Volume $\%$ change \\
	 \hline
	 Hydrostatic & 1 & 1 & 1 & 0 & 3.03 \\
	 Uniaxial along [001] & 0 & 0 & 1 & 0 & 1 \\
	 Monoclinic shear & 0 & 0 & 0 & 1 & 0 \\
	 Tetragonal shear & -0.4963 & -0.4963 & 1 & 0 & 0 \\
	 \hline
\end{tabular}
\end{center}
\label{default}
\end{table}%

In Figure 2\textit{b} for uniaxial strain along the [001] direction, results for PKAs in the [135], [100], [001] and random directions are displayed.   In Figure 2\textit{b}, it is shown that with varying strain, there is very little variation in the production of Frenkel pairs.  Some variation does exist, but all data points fall within the statistical uncertainties of the results for an unstrained system, as will be discussed below.  Therefore, application of uniaxial strain in the [001] direction, with magnitudes ranging between -2 to 2 $\%$, has no statistically significant effect on the number of stable Frenkel pairs created.  

In Figure 2\textit{c} for monoclinic shear strain, the [135], [100], [110] and random directions are displayed.  For all directions analyzed, the simulation results show no clear effect of monoclinic shear strain on damage generation.  There exists minimal variation from the unstrained system and all results fall within the statistical uncertainties of the results for an unstrained system.  

In Figure 2\textit{d}, results exploring the effect of tetragonal shear strain on defect production for PKAs along the [135], [100], [110] and random directions are displayed.  As described in Table 1, this strain state is characterized by strain with one sign along the z direction, and opposite signs along x and y, such that the volumetric strain is zero.  As the elongation in the z axis increases, the BCC unit cell is moving along the Bain path towards an FCC unit cell.  Results are shown for elongation along this path before the first inflection point in the energy versus strain curve from BCC to FCC.  From Figure 2\textit{d}, it is very difficult to determine if applied tetragonal shear strain creates a non-negligible effect on the number of stable Frenkel pairs produced.

\begin{figure}[hp]
   \centering
   \includegraphics[width=\textwidth]{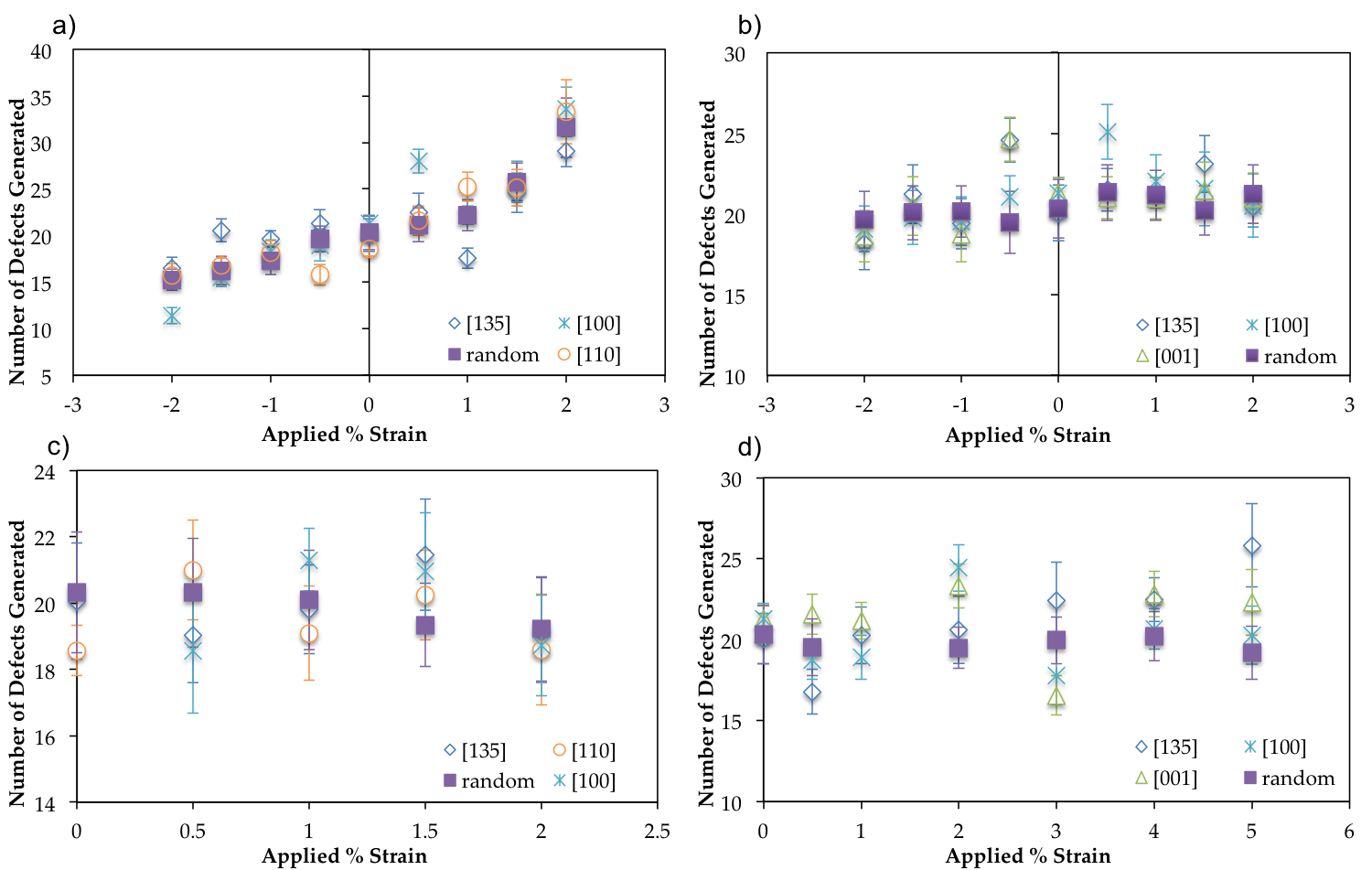} 
   \caption{Number of defects created as a function of applied (\textit{a}) hydrostatic, (\textit{b}) uniaxial, (\textit{c}) monoclinic shear and (\textit{d}) tetragonal shear strains.  PKA directions are listed in each individual figure and include the [135], [100], [110], [001] and a random set of directions.  Error bars denote twice the standard error of the mean.}
   \label{fig:example}
\end{figure}

For all types of strain, it is shown that the high symmetry directions of [100], [001] and [110] display the most variance.  This is due to the fact that a PKA moving along a high symmetry direction is highly likely to give rise to a near direct impact, potentially transporting a defect over a relatively large distance via rapid crowdion diffusion.  Thus, for a better summary of average behavior in these systems, it is most useful to look at only the results from our set of random PKA directions.  These results are displayed in Figure 3.  The error bars included are twice the standard error of the mean of the data set; i.e., they represent 95$\%$ confidence intervals in the mean value.

In Figure 3\textit{a}, for hydrostatic strain, a trend is clearly visible, with a steady increase in defects generated as the volume of the system is increased.  For small applied expansion or compression (below 1$\%$), there is not a statistically significant effect present.  The variance is approximately equal to the reported magnitude of the error bars.  However, as the expansion or compression is increased up to 2$\%$, a dramatic effect is observed.  At 2$\%$ expansion, there is a 50$\%$ increase in the number of defects generated with respect to an unstrained system.  At 2$\%$ compression, the effect is not quite as pronounced, with a 25$\%$ decrease in the number of defects generated with respect to an unstrained system.  Thus, for isotropic expansion and compression, small amounts of volume change (less than 1$\%$) lead to statistically insignificant changes in the number of defects generated.  However, larger expansion/compression leads to significant changes in the defect production.

In Figure 3\textit{b} for uniaxial strain, it is shown that very minimal changes occur to the number of defects generated by a 5 keV PKA with applied strain.  Any changes that do occur are not statistically significant for strains with magnitudes in the range of -2$\%$ to 2$\%$.

In Figure 3\textit{c} for monoclinic shear strain, minimal variation is observed as a function of applied strain.   The number of defects generated remains approximately constant as a function of applied monoclinic shear strain, leading to the same conclusion as uniaxial strains, in that there is no statistically significant impact on the number of defects generated with applied strain up to 2 $\%$.

In Figure 3\textit{d} for tetragonal shear strain, it is shown that there is a minimal amount of variation in the number of defects generated as a function of applied strain, and thus the average behavior is not impacted by applied tetragonal shear strain.  Therefore, applied tetragonal shear strain exhibits the same behavior as uniaxial and monoclinic shear strains, in that the number of defects generated remains approximately constant as a function of applied strain.

\begin{figure}[h!]
   \centering
   \includegraphics[width=\textwidth]{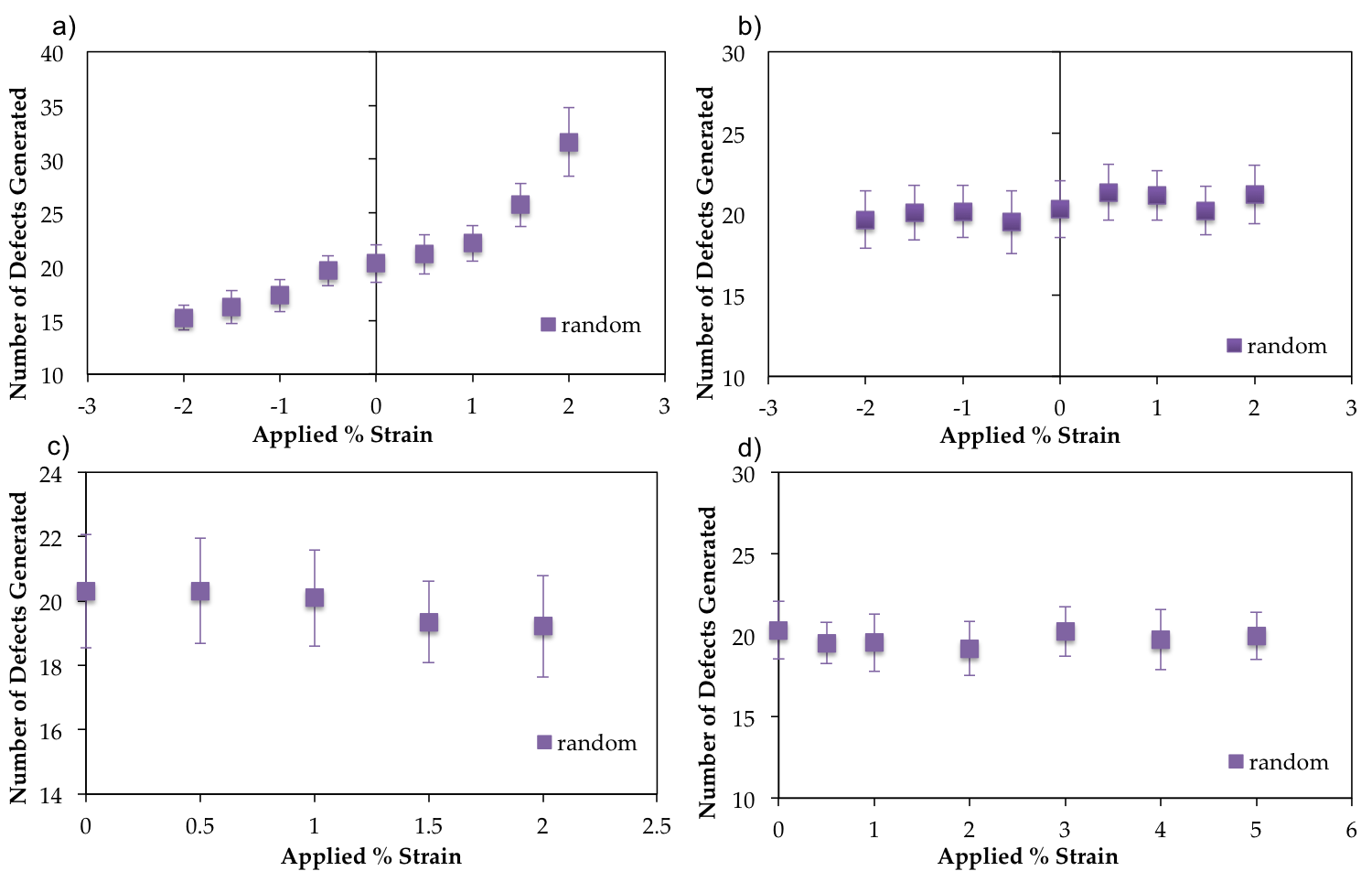} 
   \caption{Number of defects created  function of applied (\textit{a}) hydrostatic, (\textit{b}) uniaxial, (\textit{c}) monoclinic shear and (\textit{d}) tetragonal shear strains.  The results displayed are only for a set of random directions.  Error bars denote twice the standard error of the sample.}
   \label{fig:example}
\end{figure}

The results from isotropic expansion and compression investigations lead to the question of what causes the variance as a function of applied strain.  Due to the fact that the largest volume changes occurred in expansion/compression, perhaps simply volumetric changes can be the driving force for increases or decreases in defect generation.  To investigate this hypothesis, very large uniaxial strains were imposed to create volume changes approaching 2$\%$ isotropic expansion and compression.  The results of this investigation are shown in Figure 4, for uniaxial tension and compression up to 5$\%$ and 4$\%$, respectively.   In Figure 4, it is shown that from 2$\%$ compression to 2$\%$ tension, there are negligible effects of applied strain on the number of defects generated.  However, when the magnitude of the applied strain becomes more extreme, significant increases (for tension) and decreases (for compression) in the number of defects generated are observed.  For 5$\%$ applied tension, there is approximately a 40$\%$ increase in the number of defects generated.  For 4$\%$ applied compression, there is approximately a 20$\%$ decrease in the number of defects generated.  Thus, these results are comparable to the effects from isotropic expansion and compression shown in Figure 3\textit{a}, when the applied uniaxial strains lead to comparable levels of volumetric expansion or compression.  Thus, substantial increases and decreases in volume can create significant effects on the number of defects generated from a cascade in BCC Fe, whereas volume-conserving shear strains are not observed to create significant effects on the number of created defects.

\begin{figure}[h!]
   \centering
   \includegraphics[width=\textwidth]{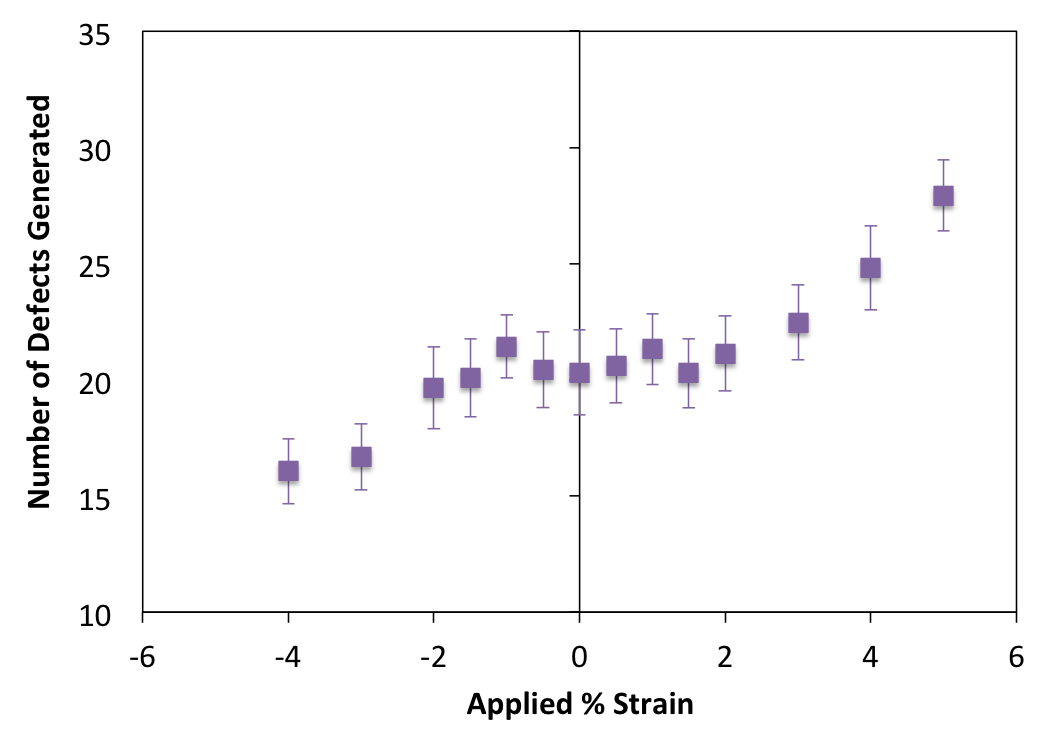} 
   \caption{Number of defects generated as a function of applied uniaxial tension/compression.  Results are displayed for up to 5$\%$ applied tension and 4$\%$ compression.  The results displayed are only for a set of random directions.  Error bars denote twice the standard error of the sample.}
   \label{fig:example}
\end{figure}

\section{Discussion}

With the observation that volume changes can affect the number of defects generated, the specific nature of how these effects take place is now investigated.  Specifically, we focus on how volume changes affect point defect formation energies, the peak number of defects generated in cascades as well as survival fractions.  This work is also performed for volume conserving strains as a comparison in order to specifically determine the effects of volume altering strains on the radiation damage behavior of BCC Fe.  

Systems with one Frenkel pair are investigated at 300 K in order to provide a direct comparison with the results on defect generation in strained systems above.  No interstitial diffusion occurs during these simulations.  Frenkel pair formation energies are calculated from equation 1

\begin{equation}
E_{form} = E_{defect} - E_{pure}
\end{equation}

where E$_{defect}$ is the energy of the system with a non-interacting interstitial/vacancy pair and E$_{pure}$ is the energy of the system with no defects.  We note that in the simulations used to compute E$_{defect}$ no interstital diffusion occurred, such that the excess energy defined in equation 1 is that of a single Frenkel defect.  Frenkel pair formation energies are displayed in Figure 5 as a function of applied \textit{a} hydrostatic, \textit{b} uniaxial, \textit{c} monoclinic shear and \textit{d} tetragonal shear strains.  Sixteen unique simulations were performed for each applied strain (eight without defects, eight with defects) to gain statistics in these systems.  Error bars in Figure 5 denote twice the standard error of the mean.  In Figure 5\textit{a} and \textit{b}, we observe significant variance in the Frenkel pair formation energy as a function of applied strain.  As applied strain becomes more positive (volume increases in the case of \textit{a} and \textit{b}), the Frenkel pair formation energy decreases.  Thus, Frenkel pair formation leads to a smaller energy change as volume increases.  The converse is true for negative strain in Figure 5\textit{a} and \textit{b}, in that a volume decrease is associated with an increase in Frenkel pair formation energy.  For the volume-conserving strains in Figure 5\textit{c} and \textit{d}, no significant changes in the Frenkel pair formation energy are observed with applied strain.  All data points possess overlapping error bars with the unstrained system.  

\begin{figure}[h!]
   \centering
   \includegraphics[width=\textwidth]{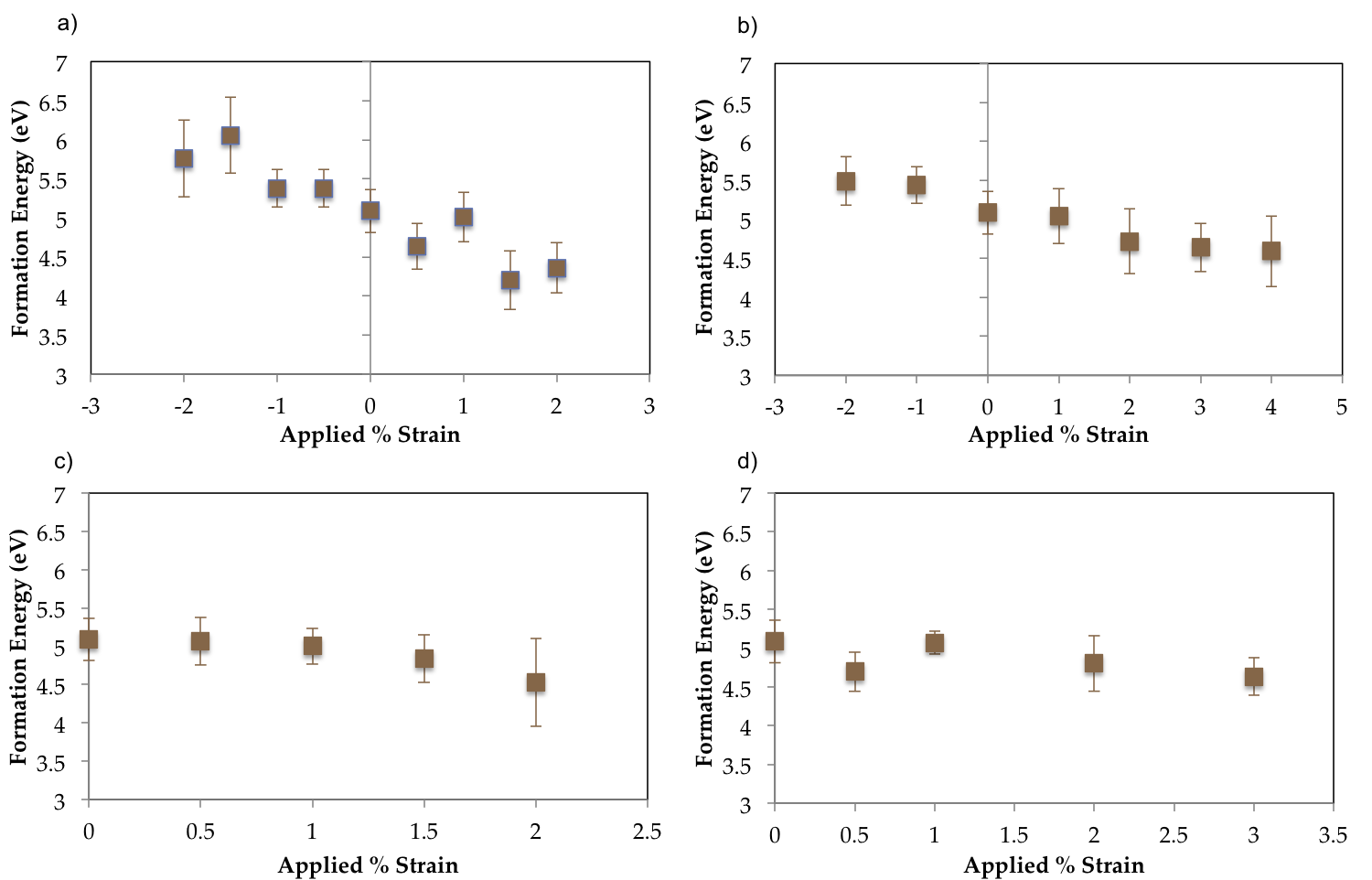} 
   \caption{Frenkel pair formation energy as a function of applied (\textit{a}) hydrostatic, (\textit{b}) uniaxial, (\textit{c}) monoclinic shear and (\textit{d}) tetragonal shear strains.  Error bars denote standard error of the sample.}
   \label{fig:example}
\end{figure}

To determine what role the variations in Frenkel pair formation energy directly play on the cascade behavior in BCC Fe, the peak number of defects created in a cascade is determined and plotted in Figure 6 as a function of applied hydrostatic, uniaxial, monoclinic shear and tetragonal shear strains.  This data is taken from the simulations of random PKA directions displayed in Figure 3.  In Figure 6\textit{a} and \textit{b}, there is a direct correlation between an increase in the volume of the system and the peak number of defects created in a displacement cascade.  For both hydrostatic and uniaxial strains, a near linear dependence is observed for the peak number of defects created in this system as a function of applied strain.  For the volume-conserving shear strains in Figure 6\textit{c} and \textit{d}, no significant changes in the peak number of defects created in a displacement cascade are observed.  From the peak number of defects in Figure 6 and the stable number of defects from Figure 3, we calculate survival fractions across all applied strains for all systems.  All survival fractions for strained systems are within 1$\%$ of the survival fraction for an unstrained system.  Therefore, applied strain produces no statistically significant variance in the defect survival fraction.  The survival fraction for all applied strains is displayed in Figure 7.  

\begin{figure}[h!]
   \centering
   \includegraphics[width=\textwidth]{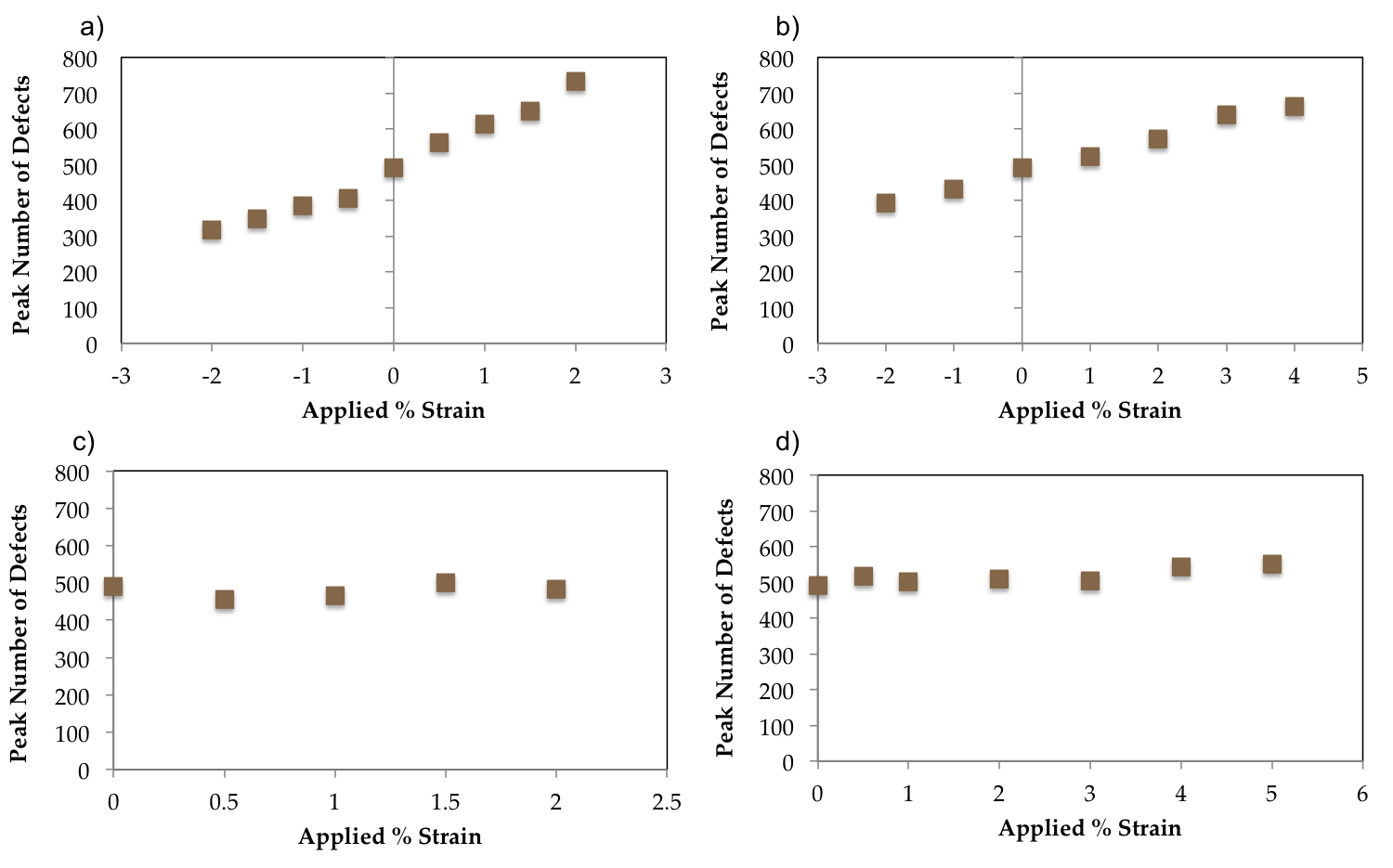} 
   \caption{Peak number of defects created  function of applied (\textit{a}) hydrostatic, (\textit{b}) uniaxial, (\textit{c}) monoclinic shear and (\textit{d}) tetragonal shear strains.}
   \label{fig:example}
\end{figure}

\begin{figure}[h!]
   \centering
   \includegraphics[width=\textwidth]{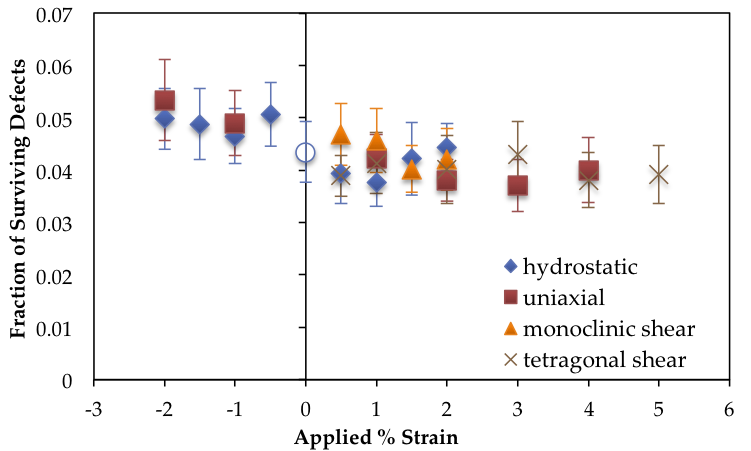} 
   \caption{Defect survival fraction as a function of applied (\textit{a}) hydrostatic, (\textit{b}) uniaxial, (\textit{c}) monoclinic shear and (\textit{d}) tetragonal shear strains.  Error bars denote standard error of the sample.}
   \label{fig:example}
\end{figure}

From these results, we can state that variations in the volume of BCC Fe create variations in the Frenkel pair formation energy, with increases in volume yielding decreases in formation energy and increases/decreases in the volume yielding conditions where it is easier/more difficult to create Frenkel pairs.  Thus, under irradiation, for a cascade of a given energy, systems with a lower Frenkel pair formation energy will exhibit a higher maximum number of created defects.  Systems with a high Frenkel pair energy will exhibit a lower maximum number of created defects.  Since volume changes produce no statistically significant changes on the survival fraction of point defects during a cascade, systems with a higher maximum number of defects created during a cascade will also exhibit a higher number of stable defects.  This is consistent with prior investigations that show that the number of defects increases with increasing PKA energy, even though the fraction of surviving defects decreases \cite{stoller1996}.  Volume-conserving strains and transformations yield no such variations in Frenkel pair formation energy and thus no such variations in the stable number of defects created via cascades in BCC Fe.  

This work suggests a direct link between formation energy and threshold displacement energy.  This is reasonable, since at 0 K the displacement energy is a combination of the formation energy of a Frenkel pair and the migration barrier to form this Frenkel pair\cite{was2007}.  Thus, it can reasonably be expected that an increase in the Frenkel pair formation energy can lead to an increase in the threshold displacement energy and a subsequent decrease in defect generation.  

To verify this interpretation, a basic investigation was undertaken to analyze the effect of strain on the displacement energy.  A BCC supercell containing 16,000 atoms is relaxed for 100 ps at 300 K in an NPT ensemble.  A strain is applied to the supercell and the strained lattice is allowed to relax for 100 ps in an NVT ensemble.  An atom near the center of the supercell is then given extra kinetic energy, with the velocity directed in a prescribed direction.  The PKA direction analyzed is a single direction from the set of 32 randomly selected directions used in the previous simulations.  The time step is set to 0.2 fs and the simulation is run for 30000 steps.  We utilize the GJF thermostat and set the damping parameter to 1 ps.  The existence of a stable Frenkel pair is determined after 6 ps via a Wigner-Seitz cell based algorithm \cite{hayward2010}.  A total of 64 independent simulations are performed on non-strained systems and 100 independent simulations are performed on strained systems in order to gain statistics.  It is important to note that since only a single direction is investigated, the results do not represent the average displacement energy, but only the displacement energy in a given direction.  Although we focused on just one randomly selected direction, we believe the results suffice to establish the overall trend related to the effect of strain on displacement energy.

In Figure 8, the probability of Frenkel pair formation as a function of PKA energy is displayed.  Red squares denote the unstrained system and a dashed trend line is fit to this data.  The blue diamonds denote the system with 2$\%$ isotropic expansion and a solid trend line is fit to this data.  A finely dashed horizontal line is overlaid at a probability of 0.5.  This is to illustrate that when the probability curve becomes greater than 0.5, it is probable to form a Frenkel pair from a given PKA energy, and thus, this value of the PKA energy is the displacement energy.  From this data, it is clear that with applied strain at a given PKA energy, the probability of Frenkel pair formation is greater.  There is also a decrease in the displacement energy with applied strain of approximately 10 eV.  This supports the previous interpretation in that applied strain affects the formation energy of Frenkel pairs and in turn affects the displacement energy.  In other words, expansion of a system will result in a decrease in the displacement energy.

\begin{figure}[h!]
   \centering
   \includegraphics[width=\textwidth]{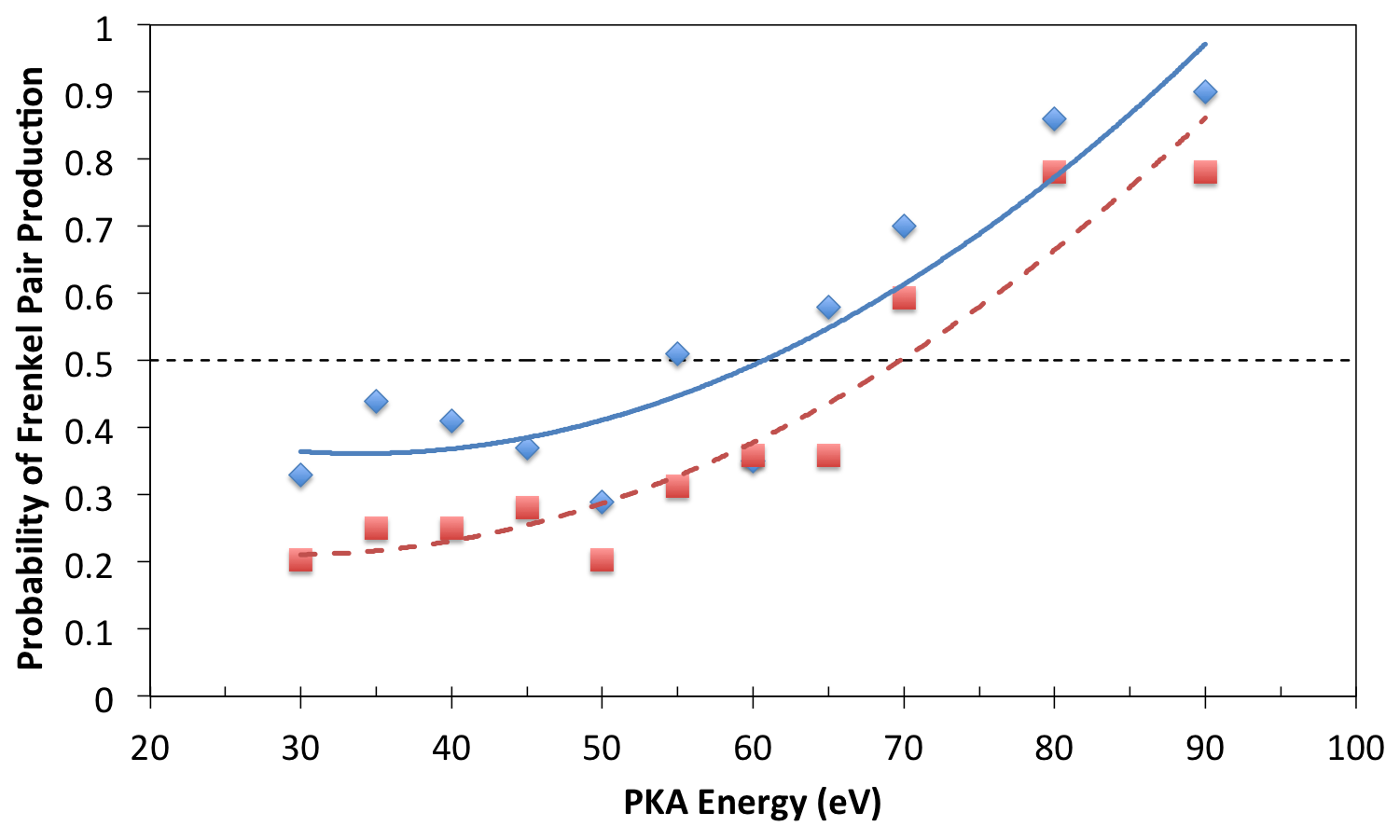} 
   \caption{Probability of Frenkel pair formation as a function of PKA energy.  Red squares denote a system without strain, blue diamonds denote a system with 2$\%$ isotropic expansion.  The energy at which the probability curve becomes greater than 0.5 is the displacement energy.}
   \label{fig:example}
\end{figure}

\FloatBarrier

\section{Conclusions}
In this study, molecular dynamics simulations were performed on pure BCC Fe to investigate the effects of applied strain on the generation of radiation damage.  The PKA directions analyzed included [135], [100], [001], [110], and finally a set of 32 randomly selected directions with a PKA energy of 5 keV at 300 K.  It was found that volume-conserving tetragonal shear and monoclinic shear strains yield no statistically significant variations in the stable number of defects created via cascades in BCC Fe.  However, isotropic expansion or compression greater than 1$\%$ or uniaxial strain greater than 2$\%$ produces a statistically significant effect on defect generation.  An increase (decrease) in the volume of the system is found to yield an increase (decrease) in the number of stable defects generated via cascades in BCC Fe.  

\section{Acknowledgement}
This work was supported by the US Department of Energy, project $\#$DE-NE0000536000.


\end{document}